\documentclass[runningheads]{llncs}
\usepackage{graphicx}
\usepackage[nolist]{acronym}
\usepackage{tikz}
\usetikzlibrary{positioning, calc, matrix, decorations.pathreplacing, shapes.geometric, fit}
\usepackage{adjustbox}
\usepackage{siunitx}
\usepackage{subcaption}
\usepackage{dirtytalk}
\usepackage{amsmath}
\usepackage[capitalise, noabbrev]{cleveref}
\usepackage[super,negative]{nth}

\usepackage{xspace}
\makeatletter
\DeclareRobustCommand\onedot{\futurelet\@let@token\@onedot}
\def\@onedot{\ifx\@let@token.\else.\null\fi\xspace}
\def\etal{\emph{et al}\onedot}

\begin{document}
\title{Domain Adversarial RetinaNet as a Reference Algorithm for the \acl{midog} Challenge}
\titlerunning{\acs{midog} Reference Algorithm}
%
\author{Frauke Wilm\inst{1}\orcidID{0000-0002-9065-0554} \and
Christian Marzahl\inst{1}\orcidID{0000-0001-9340-9873} \and
Katharina Breininger\inst{2}\orcidID{0000-0001-7600-5869} \and 
Marc Aubreville\inst{3}\orcidID{0000-0002-5294-5247}}
\authorrunning{F. Wilm et al.}
%
\institute{Pattern Recognition Lab, Computer Sciences, Friedrich-Alexander-Universit\"at, Erlangen-N\"urnberg, Germany \and
Department of Artifical Intelligence in Biomedical Engineering, Friedrich-Alexander-Universit\"at, Erlangen-N\"urnberg, Germany \and
Technische Hochschule Ingolstadt, Ingolstadt, Germany}
\maketitle              
\setcounter{footnote}{0} 
\begin{abstract}
Assessing the \acl{mc} has a known high degree of intra- and inter-rater variability. Computer-aided systems have proven to decrease this variability and reduce labeling time. These systems, however, are generally highly dependent on their training domain and show poor applicability to unseen domains. In histopathology, these domain shifts can result from various sources, including different slide scanning systems used to digitize histologic samples. The \acl{midog} challenge focused on this specific domain shift for the task of mitotic figure detection. This work presents a mitotic figure detection algorithm developed as a baseline for the challenge, based on domain adversarial training. On the challenge's test set, the algorithm scored an F$_1$~score of 0.7183. The corresponding network weights and code for implementing the network are made publicly available.   

\keywords{MIDOG \and Domain Shift  \and Mitotic Count \and Histopathology \and Object Detection}
\end{abstract}

\section{Introduction}
A well-established method of assessing tumor proliferation is the \acf{mc}~\cite{meuten2016} - a quantification of mitotic figures in a selected field of interest. Identifying mitotic figures, however, is prone to a high level of intra- and inter-observer variability~\cite{aubreville2020}. Recent work has shown that deep learning-based algorithms can guide pathologists during \ac{mc} assessment and lead to faster and more accurate results~\cite{aubreville2020}. However, these algorithmic solutions are highly domain-dependent and performance significantly decreases when applying these algorithms to data from unseen domains~\cite{lafarge2017}. In histopathology, domain shifts are often attributed to varying sample preparation or staining protocols used at different laboratories. These domain shifts and their impact on the resulting performance of an algorithm have been tackled with a wide range of strategies, e.g. stain normalization~\cite{macenko2009}, stain augmentation~\cite{tellez2018}, and domain adversarial training~\cite{lafarge2017}. Domain shifts, however, cannot only be attributed to staining variations but can also include variations induced by different slide scanners~\cite{aubreville2021}. The \acf{midog} challenge~\cite{midog_challenge}, hosted as a satellite event of the \nth{24} International Conference on \ac{miccai} 2021, addresses this topic in the form of assessing the \ac{mc} on a multi-scanner dataset. This work presents the reference algorithm developed out-of-competition as a baseline for the \ac{midog} challenge. The RetinaNet-based architecture was trained in a domain adversarial fashion and scored an F$_1$~score of 0.7183 on the final test set.

\section{Material and Methods}
The reference algorithm was developed on the official training subset of the \ac{midog} dataset~\cite{midog_data}. We did not use any additional datasets and had no access to the (preliminary) test set during method development. The algorithm is based on a publicly available implementation of RetinaNet~\cite{marzahl2020} which was extended by a domain classification path to enable domain adversarial training. 

\subsection{Dataset}
\begin{figure}[t]
\centering
\includegraphics[width=\textwidth]{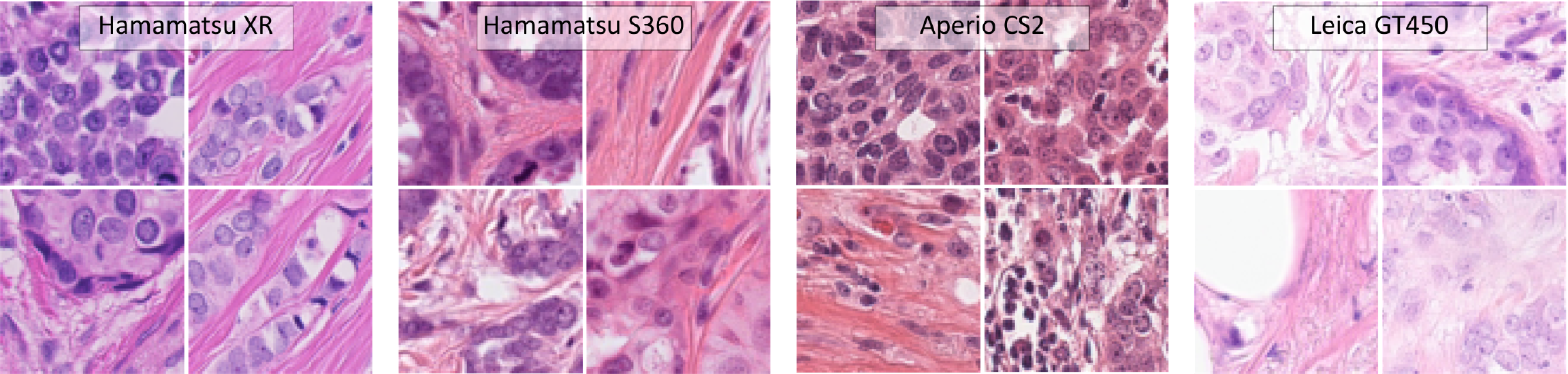}
\caption{Exemplary patches from the \acf{midog} challenge. Figure reproduced with permission from Aubreville \etal~\cite{aubreville2021}.}
\label{fig:scanners}
\end{figure}

The \ac{midog} training subset consists of \acp{wsi} from 200 human breast cancer tissue samples stained with routine \ac{he} dye. The samples were digitized with four slide scanning systems: the Hamamatsu XR, the Hamamatsu S360, the Aperio CS2, and the Leica GT450, resulting in 50 \acp{wsi} per scanner. For the slides of three scanners, a selected field of interest sized approximately \SI{2}{\milli\meter^2} (equivalent to ten high power fields) was annotated for mitotic figures and hard negative look-alikes. These annotations were collected in a multi-expert blinded set-up. Aiming to support unsupervised domain adaptation approaches, no annotations were available for the Leica GT450 so that participants could only use the images for learning a visual representation of the scanner. \Cref{fig:scanners} illustrates exemplary patches of the scanners included in the training set. 

The preliminary test set consists of five \acp{wsi} each for four slide scanning systems: the Hamamatsu XR and the Leica GT450, which already contributed to the training set, and the 3DHISTECH PANNORAMIC 1000 and the Hamamatsu RS, which were not seen during training. The scanner models of the preliminary test set, however, were undisclosed for the duration of the challenge. Participants only knew that the preliminary test set consisted of two seen and two unseen domains. This preliminary test set was used for evaluating the algorithms before submission and publishing preliminary results on a leaderboard on Grand Challenge\footnote{https://midog2021.grand-challenge.org/}. The evaluation on Grand Challenge ensured that the participants had no access to test images during method development. This restriction was also followed for developing the baseline algorithm. The final test set consists of 20 additional \acp{wsi} from the same scanners used for the preliminary test set. After the submission deadline, all algorithms were deployed once on this final test set for method comparison.

\subsection{Domain Adversarial RetinaNet}
\begin{figure}[!ht]
\centering
\input{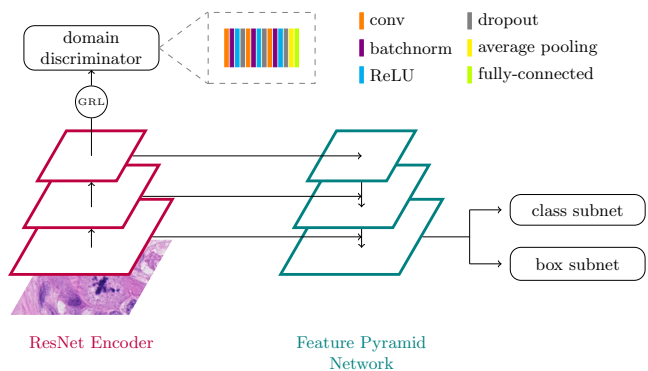}
\caption{Domain adversarial RetinaNet architecture.}
\label{fig:retinanet}
\end{figure}

For the domain adversarial training, we customized a publicly available RetinaNet implementation~\cite{marzahl2020} by adding a \ac{grl} and a domain classifier. For the encoder, we used a ResNet18 backbone pre-trained on ImageNet. For the domain discriminator, we were inspired by the work of Pasqualino \etal~\cite{pasqualino2021} and likewise chose three repetitions of a sequence of a convolutional layer, batch normalization, ReLU activation, and Dropout, followed by an adaptive average pooling and a fully connected layer. Implementation details can be obtained from our GitHub repository. We experimented with varying the number and positions of the domain classifier but ultimately decided for positioning a single discriminator at the bottleneck of the encoding branch. \Cref{fig:retinanet} schematically visualizes the modified architecture. 

\subsection{Network Training}
\label{sec:network_training}
We split our training data into 40 training and ten validation \acp{wsi} per scanner and ensured a similar distribution of high and low \ac{mc} samples in each subset. For network training, we used a patch size of 512~$\times$~512~pixels and a batch size of 12. Each batch contained three images of each scanner. To overcome class imbalance, we employed a custom patch sampling, where half of the training patches were sampled randomly from the slides and the other half was sampled in a 512-pixel radius around a randomly chosen mitotic figure. Furthermore, we performed online data augmentation with random flipping, affine transformations, and random lightning and contrast change. The loss was computed as the sum of the domain classification loss for all scanners and the bounding box regression and instance classification loss for all annotated scanners:    

{\small \begin{align*}
& \mathcal{L} =  
    \sum_{s \in S} \frac{1}{M_s} \sum_{m=1}^{M_s} \mathcal{L}_{\text{dom},m} + \beta(s) \cdot (\mathcal{L}_{\text{bb},m} + \mathcal{L}_{\text{inst},m}) & 
  \beta(s) = \begin{cases}
    0, & \text{if $s=$ GT450}.\\
    1, & \text{otherwise}. 
  \end{cases} \\
  & S: \text{set of scanners} \hspace{1cm} M: \text{samples in batch} & 
\end{align*}
}

\noindent The bounding box loss $\mathcal{L}_{\text{bb}}$ was computed as smooth L1 loss and the focal loss~\cite{lin2017} function was used for both, the instance ($\mathcal{L}_{\text{inst}}$) and the domain ($\mathcal{L}_{\text{dom}}$) classification loss. During backpropagation, the gradient was negated by the \ac{grl} and multiplied with $\alpha$, a weighting factor which was gradually increased from 0 to 1 following the exponential update scheme of Ganin \etal~\cite{ganin2015}. We trained the network with a cyclical maximal learning rate of \num{e-4} for 200 epochs until convergence. Model selection was guided by the highest performance on the validation set as well as the highest domain confusion, i.e. highest domain classification loss, to ensure domain independence of the computed features.

\subsection{Evaluation}
The training procedure described in the previous section was repeated three times and the validation slides of the three annotated scanners were used for performance assessment. To compare results across different model operating points, we constructed precision-recall curves and compared the \acp{aucpr} averaged over all three scanners for which mitotic figure annotations were available. As our final model, we selected the model with the highest mean \ac{aucpr} on the validation set and selected the operating point according to the highest mean F$_1$~score. This resulted in a mean \ac{aucpr} of 0.7551 and an F$_1$~score of 0.7369 at an operating point of 0.64 on our internal validation set. This model was submitted as a reference approach to the \ac{midog} challenge and was evaluated using a Docker-based submission system that ensured that participants of the challenge did not have access to the test images at any time during the challenge. Before the evaluation on the final test set, we ensured the sanity of the baseline algorithm by applying the model to the preliminary test set, which resulted in an F$_1$~score of 0.7401. This evaluation was run once, i.e., no hyperparameters were tuned on the preliminary test set.

For quantitative evaluation, we computed the F$_1$~score for mitosis detection on the challenge test set and compared the performance of the \say{reference approach}, trained with domain adversarial training, to a \say{weak baseline} trained without normalization or augmentation and a \say{strong baseline} trained with normalized images and the same online data augmentation methods as described in \cref{sec:network_training} but without methods for domain adaptation.


\section{Results and Discussion}
Across all test images, our weak baseline scored an F$_1$~score of 0.6279, our strong baseline an F$_1$~score of 0.6982, and our reference approach an F$_1$~score of 0.7183. Detailed results for precision, recall, and F$_1$~scores of the three models by scanner are summarized in \cref{tab:scanner_metrics}. They show that the improved F$_1$~score over the strong baseline could mainly be attributed to a higher recall, i.e. less mitotic figures were overlooked, while precision values were very similar for most scanners.  

\begin{table}[!ht]
\caption{Performance metrics per model and scanner. The Hamamatsu XR also contributed to the training set with labeled images and the Leica GT450 with unlabeled images. The other scanners were unseen during training.}
\begin{adjustbox}
{rotate=0, width=\textwidth}
\begin{tabular}{l|>{\centering}m{1.75cm}>{\centering}m{1.75cm}>{\centering}m{1.75cm}|>{\centering}m{1.75cm}>{\centering}m{1.75cm}>{\centering}m{1.75cm}|>{\centering}m{1.75cm}>{\centering}m{1.75cm}>{\centering\arraybackslash}m{1.75cm}}
    \hline
     & \multicolumn{3}{c|}{\textbf{Precision}} & \multicolumn{3}{c}{\textbf{Recall}} & \multicolumn{3}{c}{\textbf{F$_1$~score}}\\
    \cline{2-10} 
     & Weak Baseline & Strong Baseline  & Reference Approach & Weak Baseline & Strong Baseline  & Reference Approach & Weak Baseline & Strong Baseline  & Reference Approach \\
    \hline
    \textbf{Seen Domains} &&&&&&&&&\\
         XR & \textbf{0.8043}  & 0.7778 & 0.7678 & 0.7291 & 0.7586 & \textbf{0.7980} & 0.7649 & 0.7681 & \textbf{0.7826} \\
         GT450 &  \textbf{0.9016} & 0.7360 & 0.7318 & 0.2792 & \textbf{0.6650} & \textbf{0.6650} & 0.4264 & \textbf{0.6987} & 0.6968\\
    \textbf{Unseen Domains} &&&&&&&&& \\
         PANNORAMIC 1000 \hspace{0.25cm} &  0.6698 & 0.5692 & \textbf{0.6723} & 0.7172 & 0.7475 & \textbf{0.8081} & 0.6927 & 0.6463 & \textbf{0.7339}\\
         RS &  \textbf{0.6559} & 0.6417 & 0.6364 & 0.4919 & \textbf{0.6210} & \textbf{0.6210} & 0.5622 & \textbf{0.6311} & 0.6286\\
    \hline
    \textbf{All Scanners} & \textbf{0.7545} & 0.6965 & 0.7143 & 0.5377 & 0.6998 & \textbf{0.7223} & 0.6279 & 0.6982 & \textbf{0.7183}\\
    \hline
\end{tabular}
\end{adjustbox}
\label{tab:scanner_metrics}
\end{table}

\begin{figure}[!ht]
\centering
\definecolor{xr_color}{RGB}{31,119,180}
\definecolor{leica_color}{RGB}{214,39,40}
\definecolor{p1000_color}{RGB}{188,189,34}
\definecolor{rs_color}{RGB}{227,119,194}

\begin{adjustbox}
{width=0.75\textwidth}
\begin{tikzpicture}
\node[fill = xr_color](xr) at (0,0) {};
\node[right = 0.25 cm of xr.east,text depth=0pt](xr_text) {XR};
\node[fill=leica_color, right = 0.25cm of xr_text](leica){};
\node[right = 0.25 cm of leica.east](leica_text) {GT450};
\node[fill=p1000_color, right = 0.25cm of leica_text](p1000){};
\node[right = 0.25 cm of p1000.east,text depth=0pt](p1000_text) {PANNORAMIC 1000};
\node[fill=rs_color, right = 0.25cm of p1000_text](rs){};
\node[right = 0.25 cm of rs.east](rs_text) {RS};
\end{tikzpicture}
\end{adjustbox}
     \begin{subfigure}[b]{0.5\textwidth}
         \centering
         \includegraphics[width=\textwidth]{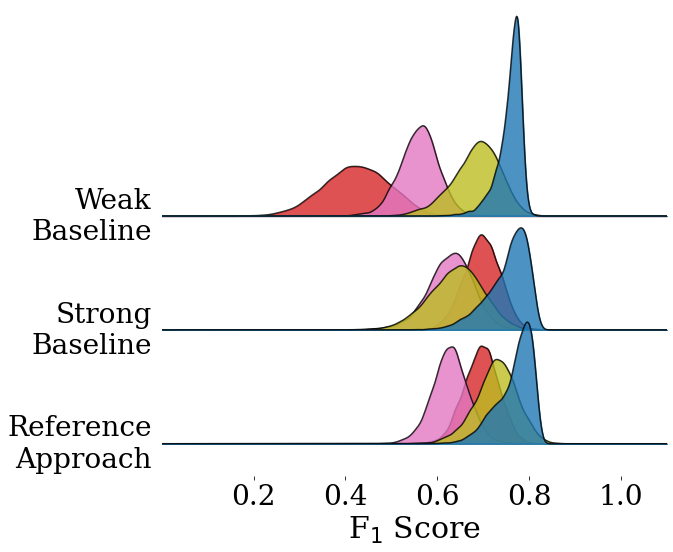}
         \caption{Bootstrapping}
         \label{fig:bootstrapping}
     \end{subfigure}\hfill
     \begin{subfigure}[b]{0.5\textwidth}
         \centering
         \includegraphics[width=\textwidth]{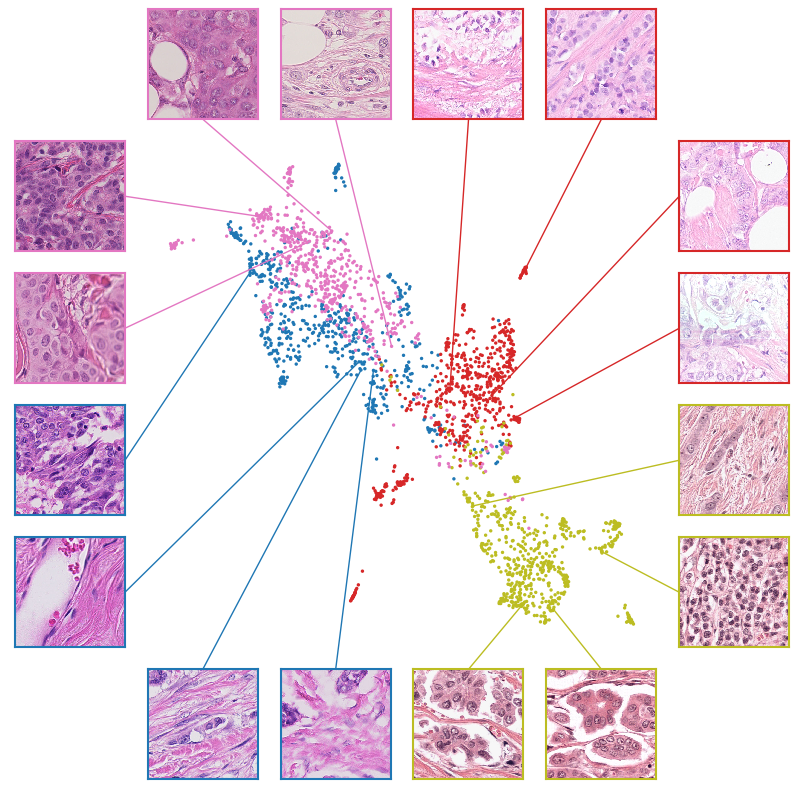}
         \caption{Weak Baseline}
         \label{fig:results_baseline}
     \end{subfigure}\hfill\\
     \begin{subfigure}[]{0.5\textwidth}
         \centering
         \includegraphics[width=\textwidth]{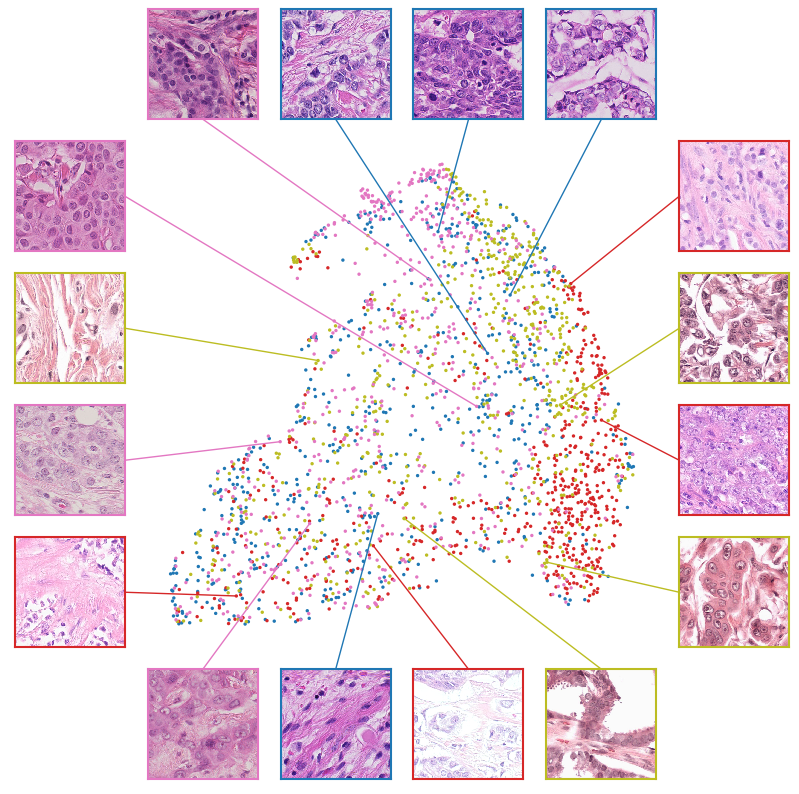}
         \caption{Strong Baseline}
         \label{fig:results_norm_aug}
     \end{subfigure}\hfill
     \begin{subfigure}[]{0.5\textwidth}
         \centering
         \includegraphics[width=\textwidth]{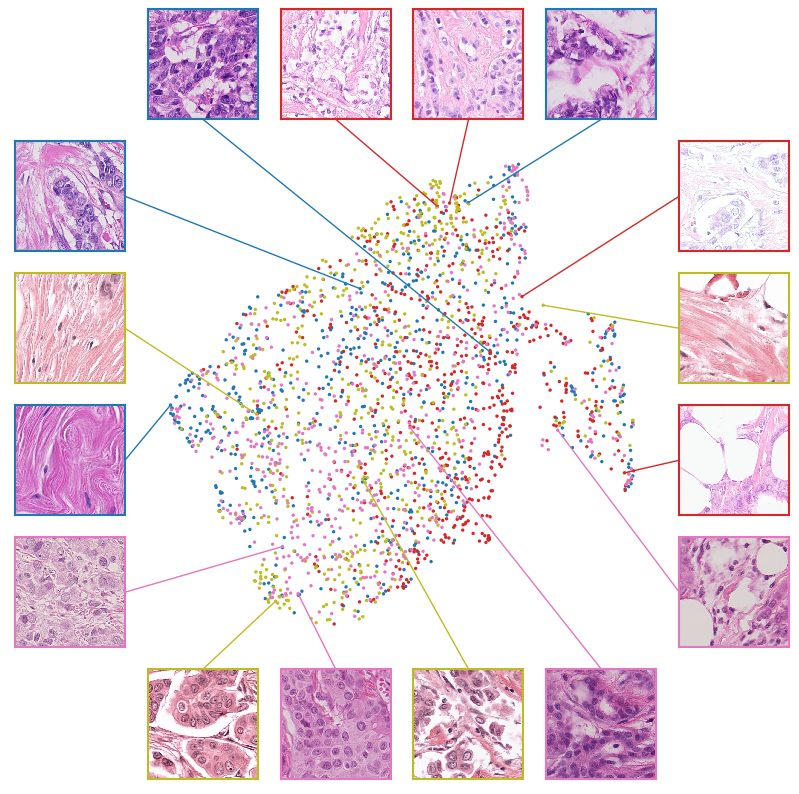}
         \caption{Reference Approach}
         \label{fig:results_da}
     \end{subfigure}\hfill
\caption{Bootstrapping and \acf{umap} plots of the evaluated models. The weak baseline was trained without any measures for normalization or augmentation and the strong baseline was trained with normalized images and online augmentations.}
\label{fig:results}
\end{figure}
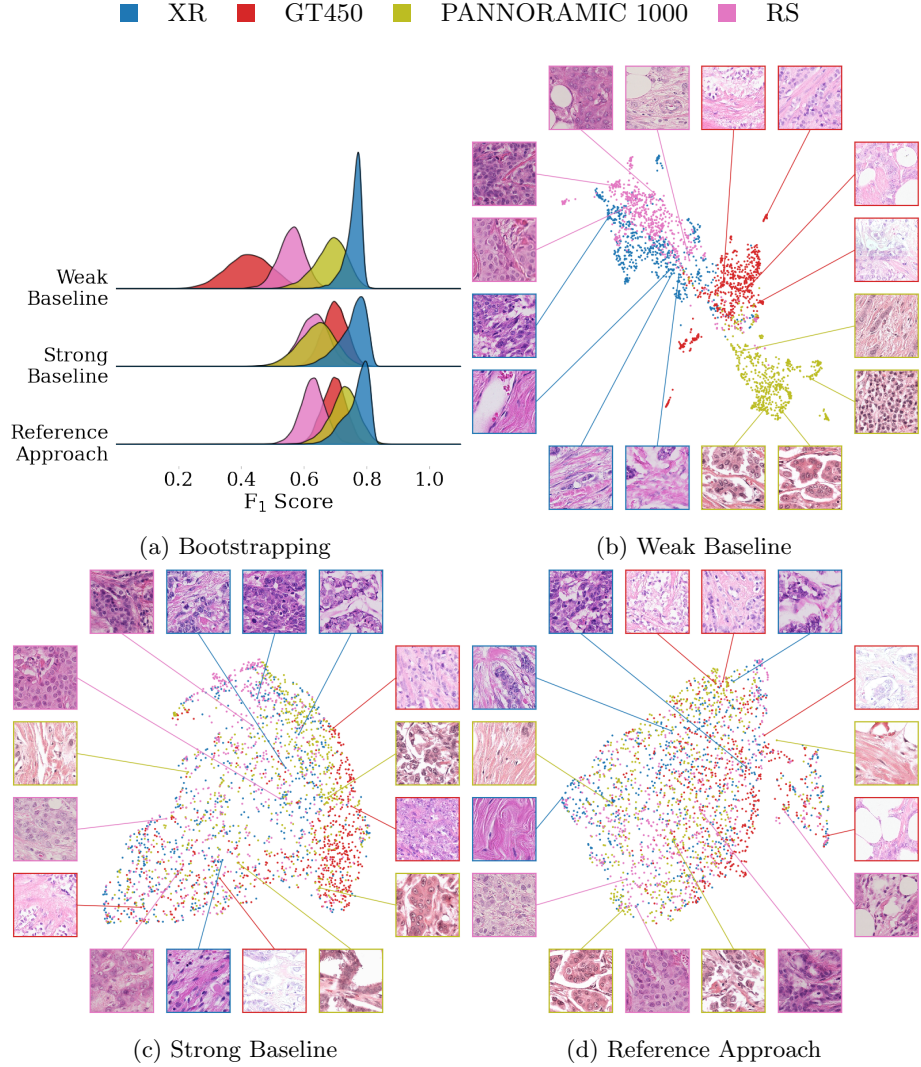

In \cref{fig:bootstrapping}, we used bootstrapping to visualize the distribution of F$_1$~scores per scanner. The results show that the weak baseline performed particularly badly for the Leica GT450 scanner with an average F$_1$~score of 0.4264 and a high variance in performance across all test slides, which becomes apparent by the wide distribution in the bootstrapping visualization. Looking at the detailed results in \cref{tab:scanner_metrics}, this was mainly attributed to a low recall, i.e. a lot of mitotic figures were overlooked. Considering the example patches of the Leica scanner shown in \cref{fig:scanners}, this result is not surprising, as the Leica scanner produces images with a much higher illumination and less contrast compared to the other scanners. Without normalization, these images can challenge the network, especially since the Leica scanner was not seen during training of the baseline models due to missing annotations and was only used for training the domain generalization component of the domain adversarial network. When comparing the strong baseline with our reference approach, the models show very similar performance for most of the scanners except for the unseen PANORAMIC 1000, where the domain adversarial training significantly increased the F$_1$~score to 0.7339 compared to an F$_1$~score of 0.6463 for the strong baseline. Furthermore, the narrower distributions of the bootstrapping in \cref{fig:results_da} indicate a lower variance in performance compared to the wider distributions of the baseline models in \cref{fig:results_baseline,fig:results_norm_aug}.

Additionally, we evaluated the models' capability for domain generalization by using \acf{umap}~\cite{mcinnes2020} plots. \ac{umap} is a dimensionality reduction technique that can be used to visualize the high dimensional feature representations within neural networks in a two-dimensional space. For our plots, we have randomly sampled 30 patches on each \ac{wsi} of the \ac{midog} test set and selected the output of the last layer of our RetinaNet encoders for visualization. The \ac{umap} plot of the reference approach is visualized in \cref{fig:results_da}. The data clustering independent of scanner domains shows that the domain adversarial training encouraged the extraction of domain-independent features. As a comparison \cref{fig:results_baseline} visualizes the \ac{umap} plot for the weak baseline. Here, the samples show a distinctive clustering according to scanner vendors. The cluster centers of the two Hamamatsu scanners are closer together, which is not surprising as they come from the same vendor and the same series (NanoZoomer). \Cref{fig:results_norm_aug} shows the \ac{umap} plot of the strong baseline. Whereas the normalization and augmentation techniques pushed the distributions closer together, the GT450 still forms a distinguishable cluster at the lower right of the feature representation. Recalling the scanner-wise model performance summarized in \cref{tab:scanner_metrics}, however, this did not impair the mitosis detection. Nevertheless, when comparing the bootstrapping visualizations in \cref{fig:results_norm_aug} and \cref{fig:results_da}, the remaining three scanners are less distinguishable in the feature representation of the domain adversarial model which seemed to have helped the mitotic figure detection for especially the unseen scanners. Interestingly, \cref{fig:results_da} shows a separated cluster on the right hand of the main cluster with patches from all scanners. A closer look at the example patches shows that these were predominantly patches with large white areas due to teared tissue or empty fat vacuoles.    

\begin{figure}[!ht]
\centering
     \begin{subfigure}[b]{0.4\textwidth}
         \centering
         \includegraphics[width=\textwidth]{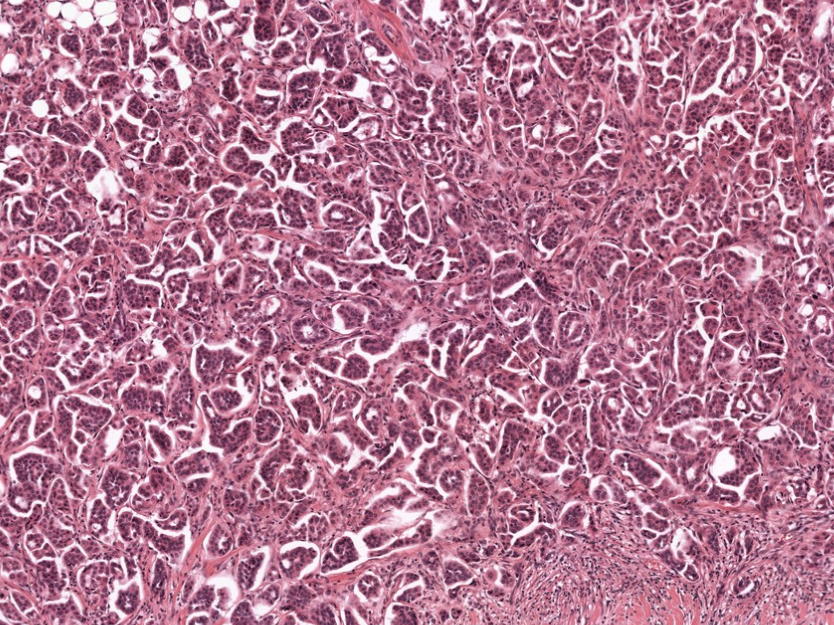}
         \caption{PANORAMIC 1000}
         \label{fig:example_p1000}
     \end{subfigure}
     \begin{subfigure}[b]{0.4\textwidth}
         \centering
         \includegraphics[width=\textwidth]{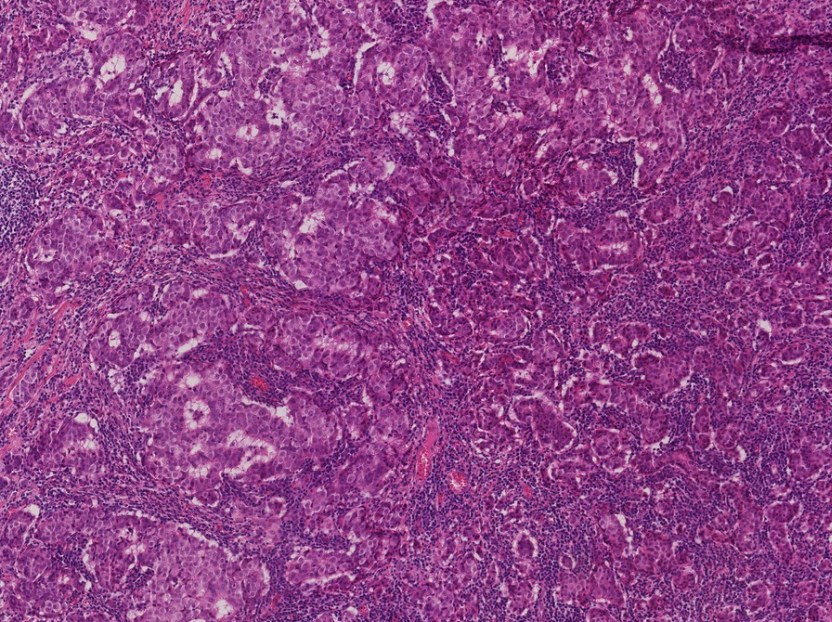}
         \caption{RS}
         \label{fig:example_rs}
    \end{subfigure}
\caption{Exemplary images where F$_1$~scores for the strong baseline and the domain adversarial varied significantly.}
\label{fig:examples}
\end{figure}

\Cref{fig:examples} shows two examples where the domain adversarial model significantly outperformed the strong baseline with F$_1$~scores of 0.8 and 0.6 for the PANORAMIC 1000 image in \cref{fig:example_p1000} and F$_1$~scores of 0.6364 and 0.4286 for the Hamamatsu RS image in \cref{fig:example_rs}. The large differences in performance could mainly be attributed to a higher number of false-positive predictions for the baseline model. Both examples show very intense staining which might not have been met with the augmentation methods used during training and thereby challenged the strong baseline model.

\section{Conclusion}
In this work, we presented our baseline algorithm for the \ac{midog} challenge, based on domain adversarial training. With an F$_1$~score of 0.7183, the algorithm is in line with previous mitotic figure algorithms trained and tested on breast cancer images from the same domain~\cite{bertram2020}. The domain adversarial training improved especially the generalization across unseen scanner domains while maintaining a similar performance on seen domains. The feature representation as \ac{umap} plots visualizes the successful extraction of domain invariant features of the proposed network. In total, 17 algorithms were submitted to the \ac{midog} challenge for evaluation on the final test set. From these, four approaches outperformed this strong but out-of-competition reference approach. The code used for implementing and training the proposed network is publicly available in our GitHub\footnote{https://github.com/DeepPathology/MIDOG} repository.

%
%
%
\bibliographystyle{splncs04}
\bibliography{mybibliography}

\begin{acronym}
\acro{mc}[MC] {mitotic count}
\acro{midog}[MIDOG]{MItosis DOmain Generalization}
\acro{miccai}[MICCAI]{Medical Image Computing and Computer Assisted Intervention}
\acro{wsi}[WSI]{Whole Slide Image}
\acro{he}[H\&E]{Hematoxylin \& Eosin}
\acro{grl}[GRL]{Gradient Reversal Layer}
\acro{aucpr}[AUCPR]{area under the precision-recall curve}
\acro{umap}[UMAP]{Uniform Manifold Approximation and Projection}
\end{acronym}

\end{document}